\documentclass[12pt]{article}
\usepackage{epsfig,amsfonts,amssymb}
\usepackage{hyperref}
\usepackage{cite}
\input epsf.sty
\usepackage{cite}

\def\ZZZ{{\hbox{ Z\kern-1.6mm Z}}}
\def\RRR{{\hbox{ R\kern-2.4mm R}}}
\def\CCC{{\hbox{ C\kern-2.0mm C}}}
\def\zzz{{\hbox{z\kern-1mm z}}}

\newcommand{\qeq}{{\hbox{=\kern-2.3mm ? \kern.5mm }}}
\renewcommand{\qeq}{=}

\newcommand{\eps}{\epsilon}

\newcommand{\ve}{\varepsilon}

\newcommand{\MM}{{\cal M}}

\newcommand{\OO}{{\cal O}}
\newcommand{\QQ}{{\cal Q}}

\newcommand{\XX}{{\cal X}}

\newcommand{\wt}{\widetilde}
\newcommand{\wh}{\widehat}

\newcommand{\be}{\begin{equation}}
\newcommand{\ee}{\end{equation}}
\newcommand{\ben}{\begin{eqnarray}\displaystyle}
\newcommand{\een}{\end{eqnarray}}

\newcommand{\refb}[1]{(\ref{#1})}
\newcommand{\p}{\partial}
\newcommand{\sectiono}[1]{\section{#1}\setcounter{equation}{0}}

\def\one{{\hbox{ 1\kern-.8mm l}}}
\def\zero{{\hbox{ 0\kern-1.5mm 0}}}

\newcommand{\bea}[1]{\begin{eqnarray}\label{#1} }
\newcommand{\eea}{\end{eqnarray}}

\newcommand{\eqref}{\refb}

\begin{document}

\begin{flushright}
\end{flushright}

\vskip 12pt

\baselineskip 24pt

\begin{center}
{\Large \bf  Covariant Action for Type IIB Supergravity}

\end{center}

\vskip .6cm
\medskip

\vspace*{4.0ex}

\baselineskip=18pt

\centerline{\large \rm Ashoke Sen}

\vspace*{4.0ex}

\centerline{\large \it Harish-Chandra Research Institute}
\centerline{\large \it  Chhatnag Road, Jhusi,
Allahabad 211019, India}

\vspace*{1.0ex}
\centerline{\small E-mail:  sen@mri.ernet.in}

\vspace*{5.0ex}

\centerline{\bf Abstract} \bigskip

Taking clues from the recent construction of the covariant action for type II and heterotic string
field theories, we construct a manifestly Lorentz
covariant action for type IIB supergravity, and
discuss its gauge fixing maintaining manifest Lorentz invariance.
 The action contains a
(non-gravitating) free 4-form field besides the usual fields of type IIB supergravity. 
This free field,
being completely decoupled from the interacting sector, has no physical consequence.

\vfill \eject

\baselineskip=18pt

\tableofcontents

\sectiono{Introduction and summary} \label{s1}

Type IIB string theory in 9+1 dimensions has a 4-form gauge potential whose 5-form
field strength obeys a self-duality constraint. As a result the theory is formulated using its
equation of motion\cite{marcus,west1,schwarz,west2} -- there is no simple Lorentz invariant
action from which the equations of motion can be
derived. Alternatively one can write down an action and supplement it with the constraint
of self-duality of the 5-form field strength. This constraint 
needs to be imposed {\it after deriving the
equations of motion from the action.}

The absence of a simple action for type IIB supergravity served as  a sort of no go theorem
for formulation of a field theory for superstrings. If a manifestly Lorentz invariant 
superstring field theory could be
formulated then by taking its low energy limit one would arrive at an action for low energy 
supergravity including type IIB supergravity. Therefore, absence of the latter would
imply absence of the former.

Recently this difficulty was circumvented and a manifestly Lorentz
invariant superstring field theory was 
formulated\cite{1508.05387}.
This theory works not only at the classical level but at the full quantum level. The
extra ingredient used in this construction was that the theory, besides containing the
usual degrees of freedom of string theory, contains a set of free fields that completely
decouple from the interacting sector, not only at the classical level but also at the
full quantum level. Given this construction one would expect that the low energy 
limit of this theory should lead to a manifestly Lorentz invariant 
action for supergravity theories, including type IIB
supergravity, at the cost of adding additional fields to the theory representing free decoupled
degrees of freedom. 

The purpose of this paper is to describe such a construction. In this we shall not try to
determine the low energy limit of the string field theory of \cite{1508.05387} directly, but use the insights
and general structure of this string field theory to guess the form of the action that 
describes type IIB supergravity. Some progress towards the study of low energy limit
of the string field theory has been achieved in \cite{okawa}. Our final result will be in the form
of an action with no additional constraints. We shall show that under suitable identification
of the field variables appearing in the new action with the field variables in the original
form of type IIB supergravity, the equations of motion derived from the new action 
reproduce both the equations of motion and the self-duality constraint on the 5-form
field strength present in type IIB supergravity. However as expected, the new formulation
has some additional degrees of freedom representing free fields
that decouple from the interacting part of the theory.

Different forms of the action for type IIB supergravity have been written down 
before. These
formulations either break manifest Lorentz 
invariance\cite{Henneaux,9304154,9701008,0605038}, 
or have infinite number
of auxiliary fields\cite{Mcclain,Wotzasek,Martin,9603031,Faddeev,9609102,9607070,9610134,9610226}, 
or have a finite number of auxiliary fields with non-polynomial  
action\cite{9509052,9611100,9707044,
9806140,9812170}, or requires going to one higher 
dimension\cite{9610234,9912086}.\footnote{Other attempts in this direction can be
found in \cite{cast1,cast2}.}
The action we construct in this paper is 9+1 dimensional,
preserves manifest Lorentz invariance, has only a finite
number of fields and  is polynomial in the fields in the absence of 
gravity.\footnote{Once
gravity is turned on the action becomes non-polynomial in the fields 
since general relativity is intrinsically
non-polynomial. Like general relativity, our action is non-polynomial in the metric
fluctuations but is polynomial in the derivatives of all fields.}
However the  general coordinate transformation acts in an
unusual fashion. This is to be expected
for two reasons. First of all in string field theory the gauge transformations look different from the
standard general coordinate transformations beyond linearized level. Therefore there is no reason to
expect that by taking its low energy limit we shall arrive at a theory with standard general coordinate
transformation rules. Second, in the standard general coordinate invariant coupling of the metric to 
other fields, in which we replace the ordinary derivatives by covariant derivatives, there are no
free fields since everything gravitates. Therefore if we are to have a field theory in which one set of
fields remain free, then the general coordinate transformation laws cannot be standard.

One way to write down a theory of 4-form fields with self-dual field strength will be to begin with a
theory of unconstrained 4-form field but arrange the interactions so that only the self-dual part
interacts with the rest of the system\cite{9610234}. In this case the anti-self-dual part would
describe a decoupled free field.
It may be possible to implement this in the full type IIB supergravity, but one has to take into
account the additional subtleties that arise from the fact that the 5-form that obey's self-duality
constraint itself depends on the interaction terms. To the best of our knowledge this has not been
carried out explicitly maintaining manifest Lorentz invariance. Here we would only like
to point out that the procedure we follow, motivated by string field theory, is different from the
one described above, In our case the extra free field that decouples also has self-dual field strength.
Furthermore it has the wrong sign kinetic term. This will be fatal in an interacting theory, but since these
extra modes describe free fields, their presence does not affect the quantization of the interacting part of
the theory.

Since the analysis of the paper is somewhat technical, let us summarize the
main results. In the usual formulation type IIB theory contains a four form gauge
potential $C^{(4)}$. The action of the theory can be written as $S_1+S_2$
where $S_2$ is independent of $C^{(4)}$, and $S_1$ has the form given in
\refb{e4.1} with the various quantities appearing in this action defined in
\refb{e2.2rep}, \refb{edefy}. 
After deriving the equations of motion using this action
we are required to impose the self-duality constraint \refb{e4.15}
on the gauge invariant 5-form
field strength.
In our formulation we replace the 4-form field $C^{(4)}$ by a 4-form field $P^{(4)}$
and an {\it independent} self-dual 5-form field $Q^{(5)}$. The action is
taken to be $S'_1+S_2$ where $S_2$ is the same action as before, and 
$S'_1$ is given in \refb{efin} with the various quantities appearing in this
expression defined in \refb{edefy}, \refb{edefeadx}, \refb{evdef}, \refb{emid}. We
find that the equations of motion derived from $S'_1+S_2$ are equivalent to
the ones derived from $S_1+S_2$ and the self-duality condition \refb{e4.15}
provided we relate the field $Q^{(5)}$ in the new formalism with the field
$C^{(4)}$ in the original formulation via eqs.\refb{e2.2rep}, \refb{e4.2}, \refb{ek7}.
The degrees of freedom associated with the field $P^{(4)}$ in the new
formalism describe (non-gravitating) 
free fields and decouple from the interacting part of the
theory. This is already apparent from the fact that $P^{(4)}$ appears in
the action \refb{efin} only in the linear and quadratic terms, but is clearer in the
gauge fixed kinetic term given in \refb{ekinetic} where the field $\bar P^{(4)}$ just
has a quadratic action and does not appear anywhere else in the action.

The rest of the paper is organized as follows. In \S\ref{s2} we use the form of the string field
theory action described in \cite{1508.05387} to guess 
the general structure of the
action for type IIB supergravity. In \S\ref{s3} we consider type IIB supergravity with the
metric fluctuations and fermion fields set to zero, and show how in this simpler setting one
can construct an action whose equations of motion reproduce the equations of motion of
type IIB supergravity. In \S\ref{s4} we include the effect of metric fluctuations as 
well as the
fermion fields and write down the general action whose equations of motion reproduce the
full set of equations of motion and self-duality constraint of type IIB supergravity.
In \S\ref{s6} we describe the general coordinate transformation laws of various fields
which take a somewhat unusual form in our description.
In \S\ref{ssuper} we describe how supersymmetry of the original
type IIB supergravity can be described as a symmetry of the
new action we have constructed in \S\ref{s4}.
In \S\ref{s6.5} we briefly discuss the Feynman rules derived from this action
in a Lorentz covariant gauge.

We expect that the formalism developed in this paper can be generalized to find
actions for other chiral theories. It will be interesting to explore if similar techniques can be
used to construct an action for the Vasiliev higher spin theories\cite{vasiliev,0304049,0503128}. If there is any limit in which the classical Vasiliev theory
emerges from classical string field theory, then the existence of an action for
the latter implies that the former must also have an action.

\sectiono{Expectation from string field theory} \label{s2}

In this section we shall review the structure of the action expected from string field
theory and describe 
how we shall implement it in the context of type IIB supergravity.

We begin by recalling some pertinent facts about the action
for superstring field theory constructed in \cite{1508.05387}. 
The theory has two sets of fields,
which we collectively denote by $\psi$ and $\wt\psi$. 
The action takes the form
\be \label{er1}
-{1\over 2} (\wt\psi, \QQ\, \XX\, \wt\psi) + (\wt\psi, \QQ \, \psi) + f(\psi)\, ,
\ee
where $(, ~)$ denotes an inner product and
$\QQ$ and $\XX$ are hermitian, mutually 
commuting, linear operators made 
of BRST charge and picture changing operators respectively. 
The details of these operators will not be
important for us. $f(\psi)$ is a non-linear function of the fields $\psi$ only,
representing interaction terms. The equations of motion for $\wt\psi$, $\psi$
derived from this action takes the form:
\be \label{er2} 
\QQ\, \XX\, \wt\psi-\QQ\,  \psi = 0\, ,
\ee
and
\be \label{er3}
\QQ\, \wt\psi + f'(\psi) = 0\, ,
\ee
where $f'(\psi)$ denotes the derivative of $f(\psi)$ with respect to various
components of $\psi$. Applying the operator $\XX$ on \refb{er3}, 
subtracting it from \refb{er2} and using the
fact that $\QQ$ and $\XX$ commute we get
\be \label{er4}
\QQ\, \psi + \XX \, f'(\psi) = 0\, .
\ee
This can be identified as the physical equations
of motion with $\psi$ containing all the physical fields. 
On the other hand \refb{er3} can now be regarded as an equation 
that determines
$\wt\psi$ in terms of $\psi$. The solution is not unique, but 
if $\wt\psi$ and $\wt\psi+\Delta\wt\psi$ represent two solutions to this
equation for a given $\psi$ then we have
\be \label{er5}
\QQ\, \Delta \, \wt\psi = 0\, 
\ee
This is a linear equation and hence 
represent free field degrees
of freedom. Furthermore, since these free field modes 
do not affect the equation for $\psi$, they 
decouple from the interacting sector described by the field
$\psi$.

The gauge symmetries of the action \refb{er1}
are generated by two sets of 
parameters collectively denoted as $\lambda$ and $\wt\lambda$. The
infinitesimal transformation laws take the form
\be \label{er6}
\delta \psi = \bar\QQ \, \lambda + \XX\, h(\psi) \, \lambda, \quad 
\delta \wt\psi = \bar\QQ\, \wt\lambda + h(\psi)\, \lambda\, ,
\ee
where $\bar\QQ$ is a field independent linear operator and
$h(\psi)$ is a linear operator acting on $\lambda$, but 
is a non-linear
function of $\psi$.

In what follows we shall use this insight to construct an action for type IIB
supergravity.  However we shall use a truncated version of this mechanism
in which we introduce the analog of the fields $\wt\psi$ 
only for the 4-form field of type IIB
supergravity. If we try to directly construct the massless field content from the
action \refb{er1} of type IIB string theory, we expect to get a doubling for 
every field
and there will also be additional auxiliary fields / gauge 
transformations etc.\cite{okawa}.

We proceed as follows. The role of
$\wt\psi$ will be played by an unconstrained 4-form field $P^{(4)}$, while the role
of $\psi$ will be played by a self-dual 5-form field $Q^{(5)}$
and all the usual
fields of type IIB supergravity  except the 4-form field. 
We shall denote these fields collectively by $M$. 
The
self-duality constraint on $Q^{(5)}$ takes the form
\be \label{e2.10}
* Q^{(5)} = Q^{(5)}\, ,
\ee
where $*$ denotes Hodge dual {\it with respect to the flat metric}.  Note that since
$Q^{(5)}$ is an independent field, this is a purely algebraic constraint.
(This will be automatic if we express $Q^{(5)}$ as a bispinor field as in type IIB
string theory.) 
The action will be
taken to be of the form\footnote{Our convention for the wedge product and $*$ 
is such that $\int dP^{(4)} \wedge * d P^{(4)} = {1\over 5!} \int
(dP^{(4)})_{\mu_1\cdots \mu_5} (dP^{(4)})^{\mu_1\cdots \mu_5}$. Therefore the
kinetic term for $P^{(4)}$ has the wrong sign. It will not affect us since 
fluctuations of $P^{(4)}$
will describe a free field.}
\be \label{e1.1}
S' = {1\over 2}  \int d P^{(4)} \wedge * d P^{(4)} - \int d P^{(4)} \wedge Q^{(5)} 
+ \wh S(Q^{(5)}, M)\, ,
\ee
where $\wh S(Q^{(5)}, M)$ will be determined by demanding that the equations of
motion derived from this action agree with those of type IIB supergravity after we
make suitable identification of the fields $(P^{(4)}, Q^{(5)})$ with the 4-form field
of type IIB supergravity in the usual formulation.
We see that as in \refb{er1},
$P^{(4)}$ appears only in the kinetic term, while $Q^{(5)}$ appears in the kinetic
term only linearly, but enters the interaction terms.
The action has gauge invariance generated by a 3-form valued parameter $\Xi^{(3)}$
\be \label{e1.15}
\delta_g P^{(4)} = d \, \Xi^{(3)}\, ,
\ee
with all other fields remaining unchanged. 
$\Xi^{(3)}$ represents a gauge transformation parameter coming from 
$\wt\lambda$. There are also other gauge transformations originating from
$\lambda$. They will be discussed later when we consider the explicit
form of $\wh S$.

Let $R^{(5)}$ denote the anti-self-dual 
5-form constructed from $M$ and $Q^{(5)}$ 
that enters the variation of 
$\wh S$ under a general variation of the fields via the relation
\be \label{e1.16}
\delta \wh S = 
-{1\over 2} 
\int  R^{(5)}\wedge \delta Q^{(5)} + \delta_M \wh S\, ,
\ee
where $\delta_M$ denotes variation with respect to all other fields labelled
by $M$. The anti-self-duality of $R^{(5)}$ is  due to the fact that
$\delta Q^{(5)}$ is self-dual and the wedge product of two self-dual 5-forms 
vanishes in 9+1 dimensions.
Then
the equations of motion for $P^{(4)}$, $Q^{(5)}$ and other fields 
derived from the action \refb{e1.1} take respectively the form:
\be \label{e1.2}
d (* dP^{(4)} - Q^{(5)}) = 0\, ,
\ee
\be\label{e1.3}
d P^{(4)} - * d P^{(4)} + R^{(5)} =0\, ,
\ee
\be \label{e1.4}
\delta_M \wh S=0\, .
\ee
Note that in writing the equation of motion \refb{e1.3} 
of $Q^{(5)}$ we have used the
fact that $Q^{(5)}$ is self-dual and that the wedge product of any self-dual tensor with another
self-dual tensor vanishes identically. 
Using
\refb{e1.3} to eliminate $* dP^{(4)}$ from the first equation we get
\be \label{e1.5}
d (Q^{(5)} - R^{(5)} ) = 0\, .
\ee
This is the analog of \refb{er4}.

We shall identify \refb{e1.5} and \refb{e1.4} as the physical equations of motion
of the theory that should reproduce the equations of motion of type IIB supergravity
once we make the correct identification of $Q^{(5)}$ with some combination
of fields of type IIB supergravity. The remaining equation \refb{e1.3} can be regarded
as the equation for $P^{(4)}$. We see from this that different solutions to 
\refb{e1.3} for given $Q^{(5)}$, $M$ 
differ from each other by free field equations of motion
\be \label{e1.6}
d (\Delta P^{(4)}) - *\, d (\Delta P^{(4)}) = 0\, .
\ee
Furthermore which solution to this equation we pick does not affect 
the physical equations encoded in \refb{e1.4}, \refb{e1.5}. Therefore the
degrees of freedom associated with $P^{(4)}$
decouple from the theory. This is also apparent from the structure of the
action -- since the interaction term does not depend on $P^{(4)}$, the 
Feynman diagrams contributing to amplitudes 
with external states associated with $Q^{(5)}$ and $M$ never
have $P^{(4)}$ propagator as internal lines.  $P^{(4)}$ only plays a role
in determining the $Q^{(5)}$-$Q^{(5)}$ propagator by inverting the off-diagonal
kinetic term in the $P^{(4)}$, $Q^{(5)}$ space after suitable gauge fixing of the
gauge symmetry \refb{e1.15}. This has been described explicitly in \S\ref{s6.5}.

\sectiono{Type IIB supergravity without gravity and fermions} \label{s3}

We begin by considering a simpler version of type IIB supergravity action where we
freeze the metric to the Minkowski metric $\eta_{\mu\nu}$ and set all the fermion
fields to zero. 
Even though this is not the full action of type IIB supergravity, this example will
illustrate how by adding free fields, we can write down manifestly 
Lorentz covariant form
for the action of interacting chiral $p$-form fields. In the next section we shall
include the effect of gravity and fermion fields.

In absence of gravity and fermions the relevant fields of type IIB supergravity
are the dilaton $\phi$ and the 2-form field $B^{(2)}$ from the NSNS sector and the 0-form field
$C^{(0)}$, 2-form field $C^{(2)}$ and the 4-form field $C^{(4)}$ in the RR sector. Let us define
\be \label{e2.1}
H^{(3)} \equiv d B^{(2)}, \quad F^{(3)} = d C^{(2)} \, ,
\ee
and\footnote{We have chosen to work in a formalism in which 
$C^{(4)}$, $F^{(5)}$ and $P^{(4)}$, $Q^{(5)}$ are invariant
under the gauge transformation associated with RR 2-form but not under the gauge 
transformation
associated with the NSNS 2-form (see \refb{e2.55},
\refb{e2.12a}). As a result we do not have 
manifest symmetry under the $SL(2,R)$ duality transformation that mixes the RR and NSNS 2-forms.
We can restore this by replacing $B^{(2)}\wedge F^{(3)}$ 
 by $(B^{(2)}\wedge F^{(3)}-C^{(2)}\wedge H^{(3)})/2$
 in all expressions in this and
the next section. 
Consequently in the gauge transformation laws of $C^{(4)}$, $P^{(4)}$, $F^{(5)}$ 
and
$Q^{(5)}$ the factors of $\lambda^{(1)}\wedge F^{(3)}$ and
$d\lambda^{(1)}\wedge F^{(3)}$ will have to be replaced respectively by
$(\lambda^{(1)}\wedge F^{(3)}-\Lambda^{(1)}\wedge H^{(3)})/2$ and
$(d\lambda^{(1)}\wedge F^{(3)}-d\Lambda^{(1)}\wedge H^{(3)})/2$.
The resulting 
formalism will have manifest $SL(2,R)$ duality symmetry with 
$C^{(4)}$, $F^{(5)}$ and $P^{(4)}$, $Q^{(5)}$ remaining invariant
under the duality rotation but now they 
will transform under the gauge transformations
associated with both the 2-form potentials. The two formalisms are related by a field
redefinition of $C^{(4)}$, $F^{(5)}$, $P^{(4)}$ and $Q^{(5)}$.}
\be \label{e2.2}
F^{(5)} \equiv d C^{(4)}, \quad \wh F^{(5)} \equiv F^{(5)} + B^{(2)} \wedge F^{(3)}\, .
\ee
Then the type IIB supergravity action is usually written as
\be \label{e2.3}
S = S_1 + S_2\, ,
\ee
where $S_2$ is a functional of all fields
other than the 4-form potential $C^{(4)}$ and 
\be \label{e2.4}
S_1 \equiv -{1\over 2} \int \wh F^{(5)} \wedge *\wh F^{(5)} + \int F^{(5)} \wedge B^{(2)} \wedge F^{(3)}
\ee
where $*$ denotes the  Hodge dual operation. The equations of motion derived from this action
have to be supplemented by the
self-duality constraint
\be \label{e2.5}
* \wh F^{(5)} = \wh F^{(5)}\, .
\ee
$S_1$ and $S_2$ are
individually invariant under the gauge transformation
\be \label{e2.55}
\delta_g B^{(2)} = d\, \lambda^{(1)}, \quad \delta_g C^{(2)} = d\, \Lambda^{(1)}, \quad 
\delta_g C^{(4)} = d\, \Lambda^{(3)} -\lambda^{(1)}\wedge F^{(3)}\, ,
\ee 
where the subscript `$g$' stands for gauge transformation. In particular
$\wh F^{(5)}$ remains invariant under these gauge transformations.

The equations of motion of $C^{(4)}$ derived from the action \refb{e2.4} takes the form
\be \label{e2.6}
d (* \wh F^{(5)} - B^{(2)} \wedge F^{(3)}) = 0\, .
\ee
This will be satisfied automatically if we use the self-duality condition \refb{e2.5} and 
the definition of $\wh F^{(5)}$ given in \refb{e2.2}. Therefore the net field equation
for $C^{(4)}$ can be summarized in the self-duality constraint \refb{e2.5} and the definition
\refb{e2.2} of $\wh F^{(5)}$. Alternatively we can treat 
$F^{(5)}$ or $\wh F^{(5)} = F^{(5)}+B^{(2)}\wedge F^{(3)}$ 
as the independent
variable and use the self-duality constraint \refb{e2.5} and the Bianchi identity
\refb{e2.6} as independent equations of motion.

The equations of motion of the rest of the fields can be expressed as
\be \label{e2.8}
\delta_M S_1 + \delta_M S_2=0\, ,
\ee
where $\delta_M$ denotes variation with respect to all other fields 
collectively denoted by $M$ {\it at fixed $F^{(5)}$}.
For our analysis
we only need to note that
\be \label{e2.9}
\delta_M S_1 = \int \left( * \wh F^{(5)} + F^{(5)}\right) \wedge \delta (B^{(2)}\wedge F^{(3)})
= \int \left( 2\wh F^{(5)} -  B^{(2)}\wedge F^{(3)}\right) \wedge \delta (B^{(2)}\wedge F^{(3)})\, ,
\ee
where in the second step we have used the self-duality constraint \refb{e2.5} and the 
relationship between $F^{(5)}$ and $\wh F^{(5)}$ given in \refb{e2.2}.

Let us now consider a different theory in which we trade in the field $C^{(4)}$ for a pair of
fields -- a 4-form field $P^{(4)}$ and an {\it independent} 5-form field $Q^{(5)}$ 
satisfying the self-duality constraint \refb{e2.10}.
We now consider the action
\be \label{e2.11}
S = S'_1 +  S_2\, ,
\ee
where  $S_2$ is the same action as what appears in \refb{e2.3}
and
\ben \label{e2.12}
S'_1 &=& {1\over 2}  \int d P^{(4)} \wedge * d P^{(4)} -
 \int d P^{(4)} \wedge Q^{(5)} 
- \int B^{(2)} \wedge F^{(3)} \wedge Q^{(5)} \nonumber \\ &&
+ {1\over 2} \int * \left(B^{(2)} \wedge F^{(3)}\right) \wedge \left(B^{(2)} \wedge F^{(3)}\right)\, .
\een
This action is invariant under the gauge transformations:
\ben \label{e2.12a}
&& \delta_g B^{(2)}=d\lambda^{(1)}, \quad \delta_g C^{(2)} = d\Lambda^{(1)}, \quad
\delta_g P^{(4)} = d \, \Xi^{(3)} - \lambda^{(1)} \wedge F^{(3)}, \nonumber \\
&& \delta_g Q^{(5)}
= - d\lambda^{(1)}\wedge F^{(3)} - * \left(  d\lambda^{(1)}\wedge F^{(3)} \right)\, .
\een
Note that we have used the same symbols $\lambda^{(1)}$ and $\Lambda^{(1)}$ 
as in the case of the previous action to indicate that these gauge transformations 
will turn out to be the same as those appearing in \refb{e2.55} once we make the
correct identification of the fields. On the other hand, 
the gauge transformation parameter 
$\Xi^{(3)}$ is {\it a priori} unrelated to $\Lambda^{(3)}$ appearing in \refb{e2.55}.

The equations of motion for $P^{(4)}$ and $Q^{(5)}$ derived from the action
\refb{e2.11}, \refb{e2.12} take the form
\be \label{e2.13}
d (* d P^{(4)} - Q^{(5)}) = 0\, ,
\ee
\be \label{e2.14}
d P^{(4)} + B^{(2)} \wedge F^{(3)} - * \left( d P^{(4)} + B^{(2)} \wedge F^{(3)} \right) = 0\, ,
\ee
respectively.
Using \refb{e2.14} to eliminate
$d*P^{(4)}$ term in \refb{e2.13}, we get
\be  \label{edq5}
dQ^{(5)} =  d (B^{(2)}\wedge F^{(3)}) - d * (B^{(2)}\wedge F^{(3)})\, .
\ee

We now claim that the theory described by the action \refb{e2.11}, \refb{e2.12}
is equivalent to that described by
the action \refb{e2.3} together with a free 4-form field with 
self-dual 5-form field 
strength, 
under the identification
\be \label{e2.14a}
\wh F^{(5)} = {1\over 2} \left[ Q^{(5)} +  B^{(2)} \wedge F^{(3)} + *\left(
B^{(2)} \wedge F^{(3)}\right)\right]\, .
\ee
For this claim to be valid the following must hold:
\begin{enumerate}
\item $\wh F^{(5)}$ defined in \refb{e2.14a} should satisfy the self-duality constraint
\refb{e2.5} and the Bianchi identity \refb{e2.6} as a consequence of
\refb{edq5}.
\item Once we make the 
identification \refb{e2.14a}, we must have
\be \label{e2.14b}
\delta_M S'_1=\delta_M S_1\, ,
\ee
so that the equations of motion for all other fields
derived from the action $S_1+S_2$ agree with those derived from the action
$S'_1+S_2$. $\delta_M S_1'$ has to be calculated at fixed $P^{(4)}$ and
$Q^{(5)}$.
\item  Given a solution to the equations of motion derived from the action
$S_1+S_2$, the identification \refb{e2.14a} should produce 
a set of solutions to the equations of motion derived from
$S'_1 + S_2$ which differ from each other by addition of plane wave 
solutions. The latter correspond to free fields and do not affect the interacting
part of the theory.
\end{enumerate}

We begin by proving the first proposition. 
$\wh F^{(5)}$ defined in \refb{e2.14a}
clearly satisfies the self-duality constraint \refb{e2.5} since $Q^{(5)}$ is
self-dual. 
Furthermore using \refb{edq5} and \refb{e2.14a} we get
\be
d \wh F^{(5)} = d (B^{(2)}\wedge F^{(3)}) = H^{(3)} \wedge F^{(3)}\, .
\ee
This agrees with \refb{e2.6}. This establishes the
first proposition.

Let us now verify the second proposition given in \refb{e2.14b}. $\delta_M S_1$
is already computed in \refb{e2.9}, so for verifying \refb{e2.14b} we need to compute
$\delta_M S'_1$. Since $P^{(4)}$ and $Q^{(5)}$ are held fixed while 
computing $\delta_M S_1'$,
we get from \refb{e2.12}:
\be
\delta_M S'_1 = 
- \int \delta (B^{(2)} \wedge F^{(3)}) \wedge Q^{(5)} 
+  \int * \left(B^{(2)} \wedge F^{(3)}\right) \wedge \delta
\left(B^{(2)} \wedge F^{(3)}\right)\, .
\ee
Using the antisymmetry of the wedge product, and \refb{e2.14a}, we can
express this as
\be
\delta_M S'_1 =  2\int \wh F^{(5)} \wedge \delta (B^{(2)} \wedge F^{(3)})
- \int (B^{(2)} \wedge F^{(3)}) \wedge
\delta (B^{(2)} \wedge F^{(3)})\, .
\ee
This agrees with $\delta_M S_1$ computed in \refb{e2.9}, thereby establishing
\refb{e2.14b}.

Finally we turn to the third proposition. Given a solution $\wh F^{(5)}$ to 
eqs.\refb{e2.5} and \refb{e2.6}, eq.\refb{e2.14a} gives us a value of $Q^{(5)}$
that solves the equations of motion \refb{edq5}. But this still
leaves open the possibility of getting different $P^{(4)}$ satisfying \refb{e2.13},
\refb{e2.14}. A particular solution to these equations is provided
by setting
\be \label{ep4c4}
P^{(4)} = C^{(4)}\, ,
\ee
where $C^{(4)}$ is related to $\wt F^{(5)}$ via \refb{e2.2}. To see this, we
note that the solution \refb{ep4c4} satisfies
\refb{e2.14} as a consequence of \refb{e2.2} and the self-duality 
condition \refb{e2.5}.  
Once \refb{e2.14} and \refb{edq5} are satisfied, 
\refb{e2.13} follows automatically. Now suppose a general solution to
\refb{e2.13}, \refb{e2.14} for $P^{(4)}$
for given $B^{(2)}$, $C^{(2)}$, $Q^{(5)}$
has the form
\be 
P^{(4)} = C^{(4)} + \wt P^{(4)}\, .
\ee
Then using \refb{e2.13}, \refb{e2.14} we get
\be \label{efree1}
d  * d \wt P^{(4)} = 0, \quad d \wt P^{(4)} - * d \wt P^{(4)} = 0\, .
\ee
Furthermore the gauge transformation generated by $\Xi^{(3)}$ acts as
\be \label{efree2}
\delta_g \wt P^{(4)} = d \, \Xi^{(3)}\, .
\ee 
Eqs.\refb{efree1} and the gauge transformation \refb{efree2} are precisely
those of a free 4-form gauge 
field with a self-duality constraint on its field strength.
Furthermore which solution of \refb{efree1} we pick does not affect the
solutions for the other fields $B^{(2)}$, $C^{(2)}$, $Q^{(5)}$ etc. which are
determined completely in terms of the solution to the equations of motion 
derived from $S_1+S_2$ via the identification \refb{e2.14a}. This establishes
the third proposition.

It is also easy to verify that the gauge transformations generated by $\lambda^{(1)}$
and $\Lambda^{(1)}$ in \refb{e2.55} agree with those given in \refb{e2.12a} 
under
the identification \refb{e2.14a}. Therefore the theory described by the action 
$S'_1+S_2$ is equivalent to the one described by the action $S_1+S_2$
and the self-duality constraint \refb{e2.5} up to addition of free fields.

Finally, note that the action \refb{e2.12} has a finite number of fields and is
polynomial in these fields. Non-polynomiality will arise when we couple this
theory to gravity, but this is an inevitable consequence of the fact that gravity
is non-polynomial.

\sectiono{Inclusion of gravity and fermions} \label{s4}

We now consider the effect of inclusion of gravity and fermions.
In this case $H^{(3)}$ and  $F^{(3)}$
are defined as in \refb{e2.1} but the definition of $\wh F^{(5)}$ is modified
to
\be \label{e2.2rep}
F^{(5)} \equiv d C^{(4)}, \quad \wh F^{(5)} = F^{(5)} 
+ Y\, , \ee
\be \label{edefy}
Y \equiv B^{(2)}\wedge F^{(3)} + \hbox{fermionic terms}
\ee
where `fermionic terms' in the definition of $Y$ describe  
5-forms constructed from the fermion bilinear. 
As in \refb{e2.3}, 
the total action is still written as
\be \label{es1s2}
S=S_1+S_2\, ,
\ee
but the action $S_1$
given in \refb{e2.4} is replaced by
\be \label{e4.1}
S_1 \equiv -{1\over 2} \int \wh F^{(5)} \wedge \star_g \wh F^{(5)} + \int F^{(5)} \wedge Y\, ,
\ee
where $\star_g$ denotes the Hodge dual operation with respect to the dynamical 
metric  $g_{\mu\nu}$. Similarly $S_2$ is covariantized with respect to the general
coordinate transformation and includes the Einstein-Hilbert term and
fermionic contribution, 
but continues to be independent of $C^{(4)}$.
The self-duality constraint \refb{e2.5} is generalized to
\be \label{e4.15}
\star_g \wh F^{(5)} = \wh F^{(5)}\, .
\ee
In order to check the internal consistency of this procedure we examine
the equations of motion for $C^{(4)}$. This takes the form
\be
d (\star_g \wh F^{(5)} - Y) = 0\, .
\ee
Using \refb{e4.15} this reduces to
\be 
d (\wh F^{(5)} - Y) = 0\, ,
\ee
which holds identically as a consequence of \refb{e2.2rep}. Therefore once we
impose the self-duality condition \refb{e4.15} and the definition \refb{e2.2rep}
of $\wh F^{(5)}$, the equation of motion for $C^{(4)}$ holds identically.

We introduce vielbein fields $\hat e_\mu{}^a$ and its inverse
$\hat E_a{}^\mu$ via
\be \label{e4.16}
\hat e_\mu{}^a \hat e_\nu{}^b \eta_{ab} = 
g_{\mu\nu}\, , \quad \hat E_a{}^\mu \hat E_b{}^\nu \eta^{ab}
=g^{\mu\nu}\, , \quad \hat e_{\mu a} = \hat e_\mu{}^b \eta_{ba}, \quad 
\hat E^{a\mu} = \eta^{ab} \hat E_b{}^\mu\, ,
\ee
and define\footnote{One cautionary comment is in order here. Often one uses the
same symbol to denote tensors under general coordinate transformation
and tensors under local 
Lorentz transformation which are related to each other by
contraction with vielbeins, {\it e.g.} $A_a = \hat E_a{}^\mu A_\mu$. We shall not use
this convention and make all factors of vielbeins explicit. For example we
have used a different symbol $\wt F$ to denote the transform of $\wh F$ to 
a tensor under local Lorentz transformation. $\wh F$ will always denote the
quantity whose components are given by the components of the right hand side
of \refb{e2.2rep} describing a tensor under general coordinate transformation.}
\be \label{e4.2}
\wt F^{(5)}_{a_1\cdots a_5} = \hat E_{a_1}{}^{\mu_1} \cdots \hat E_{a_5}{}^{\mu_5}
\wh F^{(5)}_{\mu_1\cdots \mu_5}\, .
\ee
Then the self-duality condition \refb{e4.15} on the 5-form field strength can be
reexpressed as
\be \label{e4.3} 
* \wt F^{(5)} = \wt F^{(5)}\, ,
\ee
where, as in \S\ref{s3}, $*$ now denotes the Hodge dual with respect to flat 
Minkowski metric.

In the following we shall gauge fix the local Lorentz transformation
by choosing
$\hat E^{a\mu}$ and $\hat e_{\mu a}$ to be symmetric matrices. 
The insight for this again comes from string field theory whose gauge
symmetries do not include local Lorentz transformation.
To facilitate this choice of gauge, let us express the first equation of \refb{e4.16}
in the matrix form as
\be \label{eeT}
\hat e \eta \hat e^T = g \, ,
\ee
where $\hat e$ denotes the matrix with components $\hat e_{\mu a}$.
When the metric $g_{\mu\nu}$ 
is close to $\eta_{\mu\nu}$ a solution to \refb{eeT} 
for which $\hat e_{\mu a}=\hat e_{a\mu}$ may be expressed as
\be \label{edefge}
\hat e \, \eta = (g\, \eta)^{1/2}\, ,
\ee
where in defining the square root we take the matrix that has all positive
eigenvalues. Writing $g_{\mu\nu}=\eta_{\mu\nu}+h_{\mu\nu}$,
\refb{edefge} can be written as
\be  \label{edefe}
\hat e \, \eta = (1 + h\eta)^{1/2} 
= \left(1 + {1\over 2} h\eta - {1\over 8} h\eta h\eta +\cdots\right)\, ,
\ee
so that
\be \label{ehrel}
\hat e = \left(\eta + {1\over 2} h- {1\over 8} h\eta h +\cdots\right)
\ee
is symmetric. In component this corresponds to
\be \label{edefeadx}
\hat e_{ad} = \eta_{ad} + {1\over 2} h_{ad} -{1\over 8} h_{ab} \eta^{bc} h_{cd}
+\cdots \, .
\ee
Note that in this gauge
we no longer have the distinction between the
coordinate indices $\mu,\nu,\cdots $ and the tangent space indices $a,b,\cdots$.
We shall raise and lower all indices with the flat metric $\eta$.
There is a rigid Lorentz transformation that preserves this gauge: 
under this $e_{ab}$ transforms as a covariant rank two tensor.
A general coordinate transformation must be accompanied by a
compensating local Lorentz
transformation in order to preserve this gauge.

We now introduce the following notation for operators acting on 
5-forms.  We use the indices $A, B,\cdots$ to denote the index $(a_1\cdots a_5),
(b_1\cdots b_5), \cdots$ of 5-forms. Therefore $A,B,\cdots$ each takes 
$10\choose 5$ independent 
values. However in defining sum over one of these indices
-- say $A$ -- we shall find it more convenient to define it as a sum over all values
of $a_1,\cdots a_5$, i.e.
\be 
\sum_A \equiv \sum_{a_1}\sum_{a_2}\cdots \sum_{a_5}\, .
\ee
In this notation the 5-form $\wt F^{(5)}_{a_1\cdots a_5}$
will be denoted as $\wt F^{(5)}_A$. We also introduce the following matrices
in this space:
\ben \label{evdef}
&& \zeta^{AB} = \eta^{a_1 b_1}\cdots \eta^{a_5 b_5}, \quad \zeta_{AB} = 
\eta_{a_1 b_1}\cdots \eta_{a_5 b_5}\, , \quad E^{AB} 
= \hat E^{a_1 b_1} \cdots \hat E^{a_5 b_5}\, , \nonumber \\
&&
e_{AB} = \hat e_{a_1 b_1} \cdots \hat e_{a_5 b_5}, \quad 
 \ve^{AB} 
= {1\over 5!} \eps^{a_1\cdots a_5 b_1\cdots b_5}\, ,
\een
where $\eps^{a_0\cdots a_9}$ is totally anti-symmetric in all the indices and
$\eps^{01\cdots 9}=1$.
Note that we have used the same symbol $\zeta$ for labelling a matrix with both
upper index and both lower index, but which one to use should be clear from the
expression in which it appears and the rule that an upper index can only contract
with a lower index and vice versa. For example in $\zeta e$ we have to use
$\zeta$ with upper indices while in $\zeta E$ we shall use $\zeta$ with lower index.
These matrices satisfy the identities:
\be \label{eid1}
\zeta^T=\zeta, \quad e^T=e, \quad E^T=E, \quad \ve^T=-\ve, 
\quad \ve \, \zeta\,  \ve = \zeta\, , \quad \zeta\, \zeta = I, \quad e \ve e = (-\det\hat e) \,
\zeta \, \ve\, \zeta 
\, ,
\ee
etc.\ while acting on 5-forms. 
Here $I$ denotes identity matrix and $\det \, \hat e$ is the determinant of the 
$10\times 10$ matrix $\hat e_{\mu a}$. Since all of the quantities appearing
in \refb{evdef} transform covariantly under rigid Lorentz transformation, an
action built out of these ingredients will have manifest Lorentz invariance.

The self-duality condition \refb{e4.3} on $\wt F^{(5)}$ can be expressed as
\be \label{ezf}
\zeta \, \ve \wt F^{(5)} = \wt F^{(5)}\, .
\ee
Also
in this notation \refb{e4.2} can be written as
\be \label{ek1}
\wt F^{(5)} = \zeta  E \wh F^{(5)} 
= \zeta E  (dC^{(4)} + Y)\, .
\ee
Using the fact that $e\zeta$ is the inverse matrix of $\zeta E$, we get
\be \label{ek2}
\wh F^{(5)} \equiv d C^{(4)} + Y = e\, \zeta \, \wt F^{(5)}\, .
\ee
This gives
\be \label{ek3}
d (e\zeta\wt F^{(5)} - Y) = 0\, .
\ee
We can regard $F^{(5)}= e\zeta\wt F^{(5)} - Y$ 
as independent variable instead of $C^{(4)}$,
and eqs.\refb{ezf} and \refb{ek3} as the independent
equations that determine $F^{(5)}$.

We shall now attempt to
replace the action $S_1$ by an action $S'_1$:
\be \label{espirit}
S'_1 = {1\over 2}  \int d P^{(4)} \wedge * d P^{(4)} - 
 \int d P^{(4)} \wedge Q^{(5)} + \wh S_1(Q^{(5)}, M)\, ,
\ee
and write the total action as
\be \label{etotal}
S' = S'_1 + S_2\, ,
\ee
in the
spirit of \refb{e2.11}, \refb{e2.12}. Here,
as in \refb{e1.1}, $P^{(4)}$ is an unconstrained 4-form field, $Q^{(5)}$ is a 
5-form field satisfying $Q^{(5)}=*Q^{(5)}$ and $\wh S_1$ is a 
functional of $Q^{(5)}$ and all the usual
fields of type IIB supergravity other than the 4-form field $C^{(4)}$, collectively
called $M$.  $S_2$ is the same action as what appears in \refb{es1s2}.
Our goal will be to determine $\wh S_1$ by demanding that $S'_1+S_2$
gives the same equations of motion as $S_1+S_2$ and the self-duality 
constraint \refb{ezf}, as long as we make proper
identification of fields between the two formalisms.
While doing so, we shall maintain manifest Lorentz
covariance at all stages, but invariance under general coordinate transformation
will not be manifest. 

Since $S'_1+S_2$ has the same structure as the action \refb{e1.1} with all the
$Q^{(5)}$ and $P^{(4)}$ dependence coming through $S'_1$ we have the analogs
of \refb{e1.2}, \refb{e1.3} and
\refb{e1.5} as equations of motion of $P^{(4)}$, $Q^{(5)}$:
\be \label{e1.2rep}
d (* dP^{(4)} - Q^{(5)}) = 0\, ,
\ee
\be\label{e1.3rep}
d P^{(4)} - * d P^{(4)} + R^{(5)} =0\, ,
\ee
and 
\be \label{e1.5rep}
d (Q^{(5)} - R^{(5)} ) = 0\, ,
\ee
where $R^{(5)}$ is an anti-self-dual 5-form,  defined via the equation
\be \label{e1.16rep}
\delta \wh S_1 = 
-{1\over 2} 
\int  R^{(5)}\wedge \delta Q^{(5)} + \delta_M \wh S_1\, ,
\ee
and $\delta_M$ denotes variation with respect to all other fields labelled
by $M$ at fixed $P^{(4)}$, $Q^{(5)}$.
Comparing \refb{e1.5rep} with \refb{ek3} we arrive at the identification
\be \label{ek4}
Q^{(5)} - R^{(5)} = 2(e\zeta\wt F^{(5)} - Y)\, ,
\ee
where the normalization factor of 2 on the right hand side has been chosen to
ensure compatibility between the normalization of the action \refb{e4.1} and 
\refb{espirit} (see {\it e.g.} \refb{e2.14a}).
Comparing the self-dual and anti-self-dual parts on the two sides and using
the fact that the Hodge star operation corresponds to matrix multiplication
by $\zeta \ve$ from the left, we get
\be\label{ek5}
Q^{(5)} = (1 +\zeta \ve ) e\zeta \wt F^{(5)} -
(1 + \zeta \ve) Y\, ,
\ee
and 
\be \label{ek6}
- R^{(5)} = (1 -\zeta \ve ) e\zeta \wt F^{(5)} - 
(1 - \zeta \ve) Y\, .
\ee
Our goal will be to eliminate $\wt F^{(5)}$ from these equations
to express $R^{(5)}$ as a function of $Q^{(5)}$ and
the fields $M$ 
appearing in \refb{espirit}, and then solve \refb{e1.16rep} to determine
the form of $\wh S_1$. 
For this we shall determine $\wt F^{(5)}$ in terms of $Q^{(5)}$ using
\refb{ek5} and then substitute in \refb{ek6}. Using  the
self-duality condition $\zeta \ve  \wt F^{(5)} = \wt F^{(5)}$, we can 
solve 
\refb{ek5} as
\be \label{ek7}
\wt F^{(5)} = \left\{ 1 + {1\over 2} (1 +\zeta \ve ) (e\zeta - 1) \right\}^{-1}
\left( {1\over 2} Q^{(5)}  + {1\over 2}
(1 + \zeta \ve) Y\right)\, .
\ee
Substituting this into \refb{ek6} we get
\be \label{ek8}
- R^{(5)} = {1\over 2} (1 -\zeta \ve ) e\zeta \left\{ 1 + {1\over 2} (1 +\zeta \ve ) (e\zeta - 1) \right\}^{-1}
\left( Q^{(5)}  + 
(1 + \zeta \ve) Y\right) - 
(1 - \zeta \ve) Y\, .
\ee
Using the fact that $Q^{(5)}$ and $(1+\zeta\ve)$ are  annihilated by 
$(1 - \zeta\ve)$ from the left,
we can subtract terms proportional to
$(1-\zeta\ve)Q^{(5)}$ and $(1-\zeta\ve) (1+\zeta\ve)$
from this expression.This
leads to
\be \label{ek9}
- R^{(5)} = {1\over 2} (1 -\zeta \ve ) (e\zeta - 1)  
\left\{ 1 + {1\over 2} (1 +\zeta \ve ) (e\zeta - 1) \right\}^{-1}
\left( Q^{(5)}  + %{1\over 2}
(1 + \zeta \ve) Y\right) - %{1\over 2}
(1 - \zeta \ve) Y\, .
\ee
We now note that
\be 
\int P\wedge Q= {1\over 5!}\int \ve^{AB} P_A Q_B=-{1\over 5!}
\int \ve^{AB} Q_A P_B=
-\int Q^T\ve P\, ,
\ee
where $Q^T$ denotes the transpose of $Q$ {\it multiplied by a factor of  $1/5!$.}
Using this and \refb{ek9}, \refb{e1.16rep}  gives
\ben \label{ekk1}
\delta \wh S_1 &=& -{1\over 2}  \int \delta Q^{(5)^T}  \ve (1 -\zeta \ve ) \Bigg[ 
 {1\over 2}(e\zeta - 1)  
\left\{ 1 + {1\over 2} (1 +\zeta \ve ) (e\zeta - 1) \right\}^{-1}
 Q^{(5)}  \nonumber \\ &&
+ {1\over 2}(e\zeta - 1)  
\left\{ 1 + {1\over 2} (1 +\zeta \ve ) (e\zeta - 1) \right\}^{-1}
(1 + \zeta \ve) Y - 
Y\Bigg]
+\delta_M \wh S_1\, .
\een
Our goal will be to see if we can integrate this to get $\wh S_1$.  To this end we 
define:
\be \label{emid}
\MM \equiv (\zeta-\ve) \left\{
(e\zeta - 1)  
\left( 1 + {1\over 2} (1 +\zeta \ve ) (e\zeta - 1) \right)^{-1} \zeta \right\}
\, (\ve +\zeta)
\, .
\ee
It is now easy to see using \refb{eid1} and the relation
$Q^{(5)}=\zeta \ve Q^{(5)}={1\over 2} (1+\zeta\ve) Q^{(5)}$ that we can 
express \refb{ekk1} as
\be \label{ekk2.55}
\delta \wh S_1 = {1\over 8}  \int \delta Q^{(5)^T}  \MM \,
 Q^{(5)}  
+ {1\over 2} \int \delta Q^{(5)^T}  \bigg[{1\over 2} \MM \,
 Y  - 
(\zeta-\ve) \, Y\bigg]
+\delta_M \wh S_1\, .
\ee

We can evaluate $\MM$ by first expanding the terms inside the curly bracket
on the right hand side of \refb{emid} 
in a Taylor series in $e\zeta-1$,
and then expanding each term in this series in binomial expansion.  
In any given term in this expansion containing products of $e$, $\zeta$ and
$\ve$ we can now try to reduce the number of
terms in the product using \refb{eid1} and the fact that $\ve\zeta$ acting on
$(\ve + \zeta)$ from the left gives $(\ve+\zeta)$ and $\zeta\ve$ acting on
$(\zeta-\ve)$ from the right gives $-(\zeta-\ve)$.  Using this it is easy to check that
each term in the expansion can be brought to 
$(\zeta-\ve)
(e\zeta)^n \zeta (\zeta+\ve)$ 
possibly multiplied by a power of $\det \hat e$. 
Since each of these represent a symmetric
matrix, we conclude that $\MM$ is a symmetric matrix.
Therefore the following action satisfies \refb{ekk2.55}
\be \label{ekk2.6}
\wh S_1 = {1\over 16}  \int Q^{(5)^T}  \MM \, Q^{(5)}  
+ {1\over 2} \int Q^{(5)^T}  \bigg[{1\over 2} \MM \,
Y - 
(\zeta-\ve) \, Y\bigg]
+ \wt S_1(M)\, ,
\ee
where $\wt S_1$ is independent of $Q^{(5)}$ and $P^{(4)}$
but could depend on the other fields
of the theory.

In order to determine $\wt S_1$ we have to compare $\delta_M S_1$ with
$\delta_M S'_1$. 
Recall that in computing $\delta_M S'_1$ we keep fixed $P^{(4)}$ and
$Q^{(5)}$ while in computing $\delta_M S_1$ we keep fixed $C^{(4)}$ or
equivalently $F^{(5)}$.
Now we get from \refb{espirit} and \refb{ekk2.6}:
\be
\delta_M S'_1 = \delta_M \wh S_1
= {1\over 2} \int \delta Y^T \left[
{1\over 2}\MM  - (\ve+\zeta)\right]
Q^{(5)} +\delta_M \wt S_1 + \OO(\delta \hat e)
\ee
where $\OO(\delta \hat e)$ denote terms proportional to $\delta \hat e$
-- these would come from variation of $\MM$. Note that we have
transposed the matrix sandwiched  between $Q^{(5)^T}$ and $Y$
using the symmetry of $\MM$. Using \refb{emid},  and
after some algebra, this can be expressed as
\be \label{eone}
\delta_M S'_1= - \int \delta Y^T
\ve e\zeta \left( 1 + {1\over 2} (1 +\zeta \ve ) (e\zeta - 1) \right)^{-1} Q^{(5)}
+ \delta_M \wt S_1 + \OO(\delta \hat e)\, .
\ee
On the other hand $\delta_M S_1$ can be computed from \refb{e4.1},
\refb{e2.2rep}:
\be \label{edy}
\delta_M S_1 = - \int \delta Y
\wedge (\star_g \wh F^{(5)} + F^{(5)}) +\OO(\delta \hat e)
= - \int \delta Y
\wedge \left( 2 \wh F^{(5)} - Y\right) +\OO(\delta \hat e)\, ,
\ee
where in the second step we have used the relation $\star_g \wh F^{(5)}
=\wh F^{(5)}$.
In the matrix notation this equation takes the form
\be \label{esk1}
\delta_M S_1 =- \int \delta Y^T
\ve \left( 2 \wh F^{(5)} - Y\right) +\OO(\delta \hat e)\, .
\ee
Using \refb{ek2},  \refb{ek7}, and some algebra, we arrive at
\be \label{etwo}
\delta_M S_1 =  -\int \delta Y^T
\ve\, e\, \zeta \left( 1 + {1\over 2} (1 +\zeta \ve ) (e\zeta - 1) \right)^{-1} Q^{(5)}
%\nonumber \\ &&
- \int \delta Y^T \zeta \, Y
+ {1\over 2} \int \delta Y^T  \MM  \,
Y  +\OO(\delta \hat e)\, .
\ee
Comparing \refb{eone} and \refb{etwo} we get
\be
\delta_M \wt S_1 = - \int \delta Y^T 
\zeta \, Y
+{1\over 2} \int \delta Y^T  \MM \, Y   +\OO(\delta \hat e)\, ,
\ee
and hence 
\be \label{ethree}
\wt S_1 = -{1\over 2} \int Y^T 
\zeta \, Y
+{1\over 4} \int Y^T  \MM  \,
Y + \cdots\, ,
\ee
where $\cdots$ now denotes some functional of $\hat e_{a\mu}$ only. 
However such a term, under a variation of $\hat e_{a\mu}$, will give a 
non-vanishing contribution to 
$\delta_M S'_1$ even when $Q^{(5)}$ and $Y$ vanish.
It is easy to see from \refb{e4.1} that $\delta_M S_1$ does not
have such terms. Therefore if we want the equality of
$\delta_M S'_1$ and $\delta_M S_1$ to hold even for variation with
respect to $\hat e_{a\mu}$, then 
the $\cdots$ terms in \refb{ethree} must vanish.
Therefore we get from \refb{espirit}, \refb{ekk2.6}, \refb{ethree}
\ben \label{efin}
S'_1 &=& {1\over 2}  \int d P^{(4)} \wedge * d P^{(4)} 
- \int d P^{(4)} \wedge Q^{(5)} %\nonumber \\ && 
+ {1\over 16}  \int Q^{(5)^T}     \MM \,
 Q^{(5)}  \nonumber \\ &&
+ {1\over 2} \int Q^{(5)^T}  \bigg[{1\over 2} \MM \,
Y - (\zeta-\ve) \,
Y\bigg] %\nonumber \\ &&
-{1\over 2}\int Y^T 
\zeta \,Y
+{1\over 4} \int Y^T  \MM\,  Y \, . % \nonumber \\
\een
This is what should replace the action \refb{e4.1} in this formulation.
Note that unlike in the case of the action \refb{e4.1}, where a 
self-duality constraint has to be imposed after deriving the equations 
of motion, there is no such additional constraint 
for the action \refb{efin}.

In order to verify that the classical equations of motion derived from 
\refb{efin} are equivalent to the usual equations of motion of type IIB string
theory, we also need to check that the variation of $S'_1$ with respect
to $\hat e_{ab}$ at fixed $P^{(4)}$, $Q^{(5)}$ 
agrees with the variation of $S_1$ with respect to 
$\hat e_{ab}$ at fixed $F^{(5)}$.
In making this comparison we can ignore possible dependence
on $\hat e_{ab}$ entering through $Y$ since we have already ensured that
the terms involving $\delta Y$ agree between $\delta_M S_1$ and $\delta_M
S'_1$.
Let us denote by $\delta_e$ the variation 
with respect to $\hat e_{ab}$  at fixed $Y$, $P^{(4)}$, $Q^{(5)}$ while .acting
on $S'_1$ and fixed $Y$, $F^{(5)}$ while acting on $S_1$. We need to
show the equality of $\delta_e S_1$ and $\delta_e S'_1$.
Now from \refb{efin} we have
\be
\delta_e S'_1 = {1\over 16}  \int Q^{(5)^T}     \delta\MM \,
 Q^{(5)} + {1\over 4} \int Q^{(5)^T}   \delta\MM \,
Y +{1\over 4} \int Y^T \delta \MM\,  Y .
\ee
Using \refb{ek7} and the results $(1-\ve\zeta)\delta\MM=2\,
\delta\MM=\delta\MM
(1+\zeta\ve)$ that follows from \refb{emid}, we can express this as
\be \label{es1pre}
\delta_e S'_1 = {1\over 4} \int \wt F^{(5)T} \left\{ 1 
+ {1\over 2} (1 +\zeta \ve ) (e\zeta - 1) \right\}^T \delta\MM \left\{ 1 + {1\over 2} (1 +\zeta \ve ) (e\zeta - 1) \right\}
\wt F^{(5)}\, .
\ee
From \refb{emid} we get
\be 
\delta \MM = (\zeta -\ve)\, \left( 1 + {1\over 2} (e\, 
\zeta-1)(1+\zeta\, \ve)\right)^{-1}\, \delta e \, \zeta \, 
\left( 1 + {1\over 2} (1+\zeta\, \ve) (e\, \zeta - 1)\right)^{-1} \, (1+\zeta\, \ve)\, .
\ee
Using this, and the result $(1+\zeta\, \ve)\wt F^{(5)}= 2 \wt F^{(5)}$, we get
\be \label{es1}
\delta_e S'_1 = {1\over 2} \int
\wt F^{(5)T} \left\{ 1 
+ {1\over 2} (1 +\zeta \ve ) (e\zeta - 1) \right\}^T
(\zeta -\ve)\, \left( 1 + {1\over 2} (e\, 
\zeta-1)(1+\zeta\, \ve)\right)^{-1}\, \delta e \, \zeta \, \wt F^{(5)}\, .
\ee

Let us now turn to the computation of $\delta_e S_1$.
To make the metric dependence of 
\refb{e4.1} manifest we introduce the matrix notation:
\be 
G^{AB} = g^{a_1b_1} \cdots g^{a_5 b_5} \quad \hbox{for} \quad
A=(a_1,\cdots a_5), \quad B=(b_1,\cdots b_5)\, ,
\ee
and express \refb{e4.1} as
\be
S_1 = -{1\over 2\times 5!} \int \sqrt{-\det g} \, \wh F^{(5)}_A  G^{AB} \wh F^{(5)}_B 
+ {1\over 5!}  \int F^{(5)}_A \ve^{AB} \, Y_B\, .
\ee
This gives
\be
\delta_e S_1 = -{1\over 2 \times 5!} \int (\delta \sqrt{-\det g}) \,
\wh F^{(5)}_A  G^{AB} \wh F^{(5)}_B -
{1\over 2\times 5!} \int \sqrt{-\det g} \, \wh F^{(5)}_A  \delta G^{AB} \wh F^{(5)}_B \,.
\ee
The first term vanishes due to the self-duality constraint \refb{e4.15} on
$\wh F^{(5)}$. The second term can be simplified using \refb{ek2} and
\be 
G = E \zeta E\, , \quad \sqrt{-\det g} = -\det  \hat e\, .
\ee
This gives, recalling that the definition of $\wt F^{(5)T}$ includes a transpose and
a multiplicative factor of $1/5!$:
\be 
\delta_e S_1 =  -
{1\over 2} \int (-\det  \hat e) \, \wt F^{(5)T} \, \zeta \, e\, 
(\delta E \, \zeta \, E + E\, \zeta\,
\delta E) \, e \, \zeta \, \wt F^{(5)}\, .
\ee
Since $E=e^{-1}$ we have $\delta E = - e^{-1} \delta e e^{-1}$. Using this
we can simplify this equation as
\be \label{e4.57}
\delta_e S_1 = {1\over 2} \int (-\det  \hat e) \, \wt F^{(5)T} (\zeta\, \delta e\, e^{-1}
+ e^{-1} \, \delta e\, \zeta) \wt F^{(5)}
= \int (-\det  \hat e) \, \wt F^{(5)T} \, e^{-1} \, \delta e\, \zeta\, \wt F^{(5)}
\, ,
\ee
where in the second step we have replaced the first term in the middle expression
by its transpose. 
Using \refb{eid1} and \refb{ezf} we can write
\be 
\wt F^{(5)T} = - \wt F^{(5)T} \, \ve\, \zeta =- \wt F^{(5)T} \, \zeta\, \zeta \,
\ve\, \zeta =
 - (-\det  \hat e)^{-1} \, \wt F^{(5)T} \, \zeta \, e \, \ve \, e\, .
\ee
Substituting this into \refb{e4.57} we get
\be \label{ess2}
\delta_e S_1 = - \int  \wt F^{(5)T} \zeta \, e\, \ve\, \delta e\, \zeta\, \wt F^{(5)} \, .
\ee

From \refb{es1} and \refb{ess2} we now get
\ben
&& \delta_e  S'_1 - \delta_e S_1 \nonumber \\
&=&  {1\over 2} \int
\wt F^{(5)T} \left[ \left\{ 1 
+ {1\over 2} (1 +\zeta \ve ) (e\zeta - 1) \right\}^T
(\zeta -\ve) + 2 \, \zeta \, e\, \ve \, \left( 1 + {1\over 2} (e\, 
\zeta-1)(1+\zeta\, \ve)\right)\right] \nonumber \\ && 
\left( 1 + {1\over 2} (e\, 
\zeta-1)(1+\zeta\, \ve)\right)^{-1}\, \delta e \, \zeta \, \wt F^{(5)}\, .
\een
Straightforward manipulation of the expression inside the square bracket
using the self-duality of $\wt F^{(5)}$ gives
\be \label{efincan}
\delta_e  S'_1 - \delta_e S_1 = 0\, .
\ee
This shows complete equivalence between the equations of motion derived from
$S_1+S_2$ and $ S'_1+S_2$.

Given the action $S_1'+S_2$, one can formally quantize this using
Batalin-Vilkovisky (BV) formalism following the same route 
as for the string field theory action of 
\cite{1508.05387}, since the structure of gauge transformations in the two
theories are similar. However
this theory will suffer from the usual ultraviolet divergences of superstring theory. Therefore
the full quantization of the theory will require using the full string field theory. For this
reason,
the utility of this action lies not in the fact that 
we can use it to quantize type IIB supergravity, but in that
it is through action of this type that one can make a direct link between
superstring field theory -- needed for a systematic quantization of superstring theory --
and supergravity describing the low energy dynamics of the theory. 
This construction also throws light on an apparent puzzle -- it has been known since early days of
string theory that the
RR vertex operators in the canonical $(-1/2, -1/2)$ picture couple directly to the RR field
strengths instead of RR gauge fields, while supergravity theories, including type IIA supergravity
which has an action, are naturally formulated in terms of the gauge fields. The
formulation of type IIB supergravity presented here illustrates how supergravity theories 
can be formulated directly in terms of field strengths. This also tells us that the 
version of supergravity that will emerge naturally from string field theory should
involve a formulation in which the other RR fields are also described in terms of their field strengths.
Such a formulation can be easily obtained by generalizing the construction described here, although this
is not necessary for being able to write down the action.

\sectiono{General coordinate transformation} \label{s6}

Even though our action is not manifestly invariant under 
general coordinate transformation, since the
equations of motion are equivalent to the equations of motion of type IIB 
supergravity,
the formalism has hidden general coordinate invariance. In this section we 
shall determine the general coordinate transformation laws for various fields in
our formalism.

First of all,  for all the fields other than $P^{(4)}$ and $Q^{(5)}$, collectively called $M$ 
in \refb{e1.1},  the general coordinate
transformation rules are the same as in the original formulation of type IIB supergravity
since these fields are in one to one correspondence in the two formalisms. This
includes also the vielbein fields, but we should keep in mind that general
coordinate transformations have to be accompanied by a local Lorentz transformation
to keep the vielbein symmetric. 
Since under combined infinitesimal general coordinate and local Lorentz
transformations generated by the parameters $\xi^a$ and $\omega_{ab}
=-\omega_{ba}$, we have
\be 
\delta \hat e_{ab} = \p_a \xi^c \, \hat e_{cb} + \xi^c \, \p_c \hat e_{ab} 
+\hat e_{ac} \, \eta^{cd} \, \omega_{db}\, ,
\ee
the requirement that $\delta \hat e_{ab}$ remains symmetric determines
$\omega_{ab}$ in terms of $\xi^a$ via the relations:
\be 
\hat e_{ac}\, \eta^{cd}\, \omega_{db} - \hat e_{bc} \, \eta^{cd} \, \omega_{da}
= \p_b \xi^c \, \hat e_{ca} -  \p_a \xi^c \, \hat e_{cb} \, .
\ee
This compensating local Lorentz transformation must also act on the  fermions
under a general coordinate transformation.

For determining the transformation laws of $P^{(4)}$ and $Q^{(5)}$ we again draw our
insight from the structure of gauge transformations in superstring field theory given 
in \refb{er6}. In the truncated version of the theory in which the only field coming
from $\wt\psi$ is $P^{(4)}$, the gauge transformation parameters are also truncated
with $\wt\lambda$ giving only the 3-form gauge transformation parameter 
$\Xi^{(3)}$ that generates $P^{(4)}\to P^{(4)}+d\, \Xi^{(3)}$ transformation,
and $\lambda$ containing all other gauge transformation
parameters including general coordinate transformation. If we denote by 
$\delta_\xi$ the general coordinate transformation with infinitesimal parameter
$\xi$ then it follows from \refb{er6} that 
$\delta_\xi P^{(4)}$ and $\delta_\xi Q^{(5)}$ 
will be a function of
the fields coming from $\psi$. This includes
$Q^{(5)}$ and other fields collectively denoted by $M$ in \refb{e1.1} 
but does not include
$P^{(4)}$. This is clearly an unusual transformation law for $P^{(4)}$ since
even the usual term $\xi^a \p_a P^{(4)}$ will not be present in the
transformation of $P^{(4)}$.\footnote{Unusual form of general coordinate 
transformation laws also appear in 
double field theories\cite{9302036,9305073,9308133,0904.4664}.}

We now note the following:
\begin{enumerate}
\item We have seen in \S\ref{s4} that the variation $\delta_M$ 
of $S_1$ with respect to all fields at
fixed $F^{(5)}$ agrees with the variation $\delta_M$ of $ S'_1$ with respect
to all fields at fixed $P^{(4)}$ and $Q^{(5)}$. As a special case, this also applies 
to variations induced by general
coordinate transformation.
\item $S_1$ given in \refb{e4.1}
is manifestly invariant under general coordinate transformation
{\it if we regard $F^{(5)}$ as an independent 5-form field variable}
and $\wh F^{(5)}$ to be given by $F^{(5)}+Y$. 
\item Therefore if we can ensure that the variation of $ S'_1$ under the general coordinate 
transformation of $P^{(4)}$ and $Q^{(5)}$ agrees with the variation of $S_1$ under the
general coordinate transformation of $F^{(5)}$ 
then we would have
proved the general coordinate invariance of $ S'_1$. 
Denoting 
by the symbol $\delta'_\xi$ the transformation
of the action induced by the general coordinate transformation of $F^{(5)}$, 
$P^{(4)}$
and $Q^{(5)}$ only, the requirement given above translates to
\be \label{equal}
\delta'_\xi  S'_1 = \delta'_\xi S_1\, .
\ee
\item Since $S_2$ is manifestly invariant under general coordinate transformation, this will
also prove the general coordinate invariance of $ S'_1+S_2$.
\end{enumerate}
Note the emphasis on the fact that in computing $\delta'_\xi S_1$
we
need to regard $F^{(5)}$ as an independent field variable.
The reason for this is as follows.
As is well known, while
checking symmetries of the action under a given transformation 
we cannot use 
equations of motion. For $ S'_1$ this translates to the statement that we
should not use eqs.\refb{e1.2rep}, \refb{e1.3rep} and \refb{e1.5rep}. However
we are allowed to use the self-duality of $Q^{(5)}$  
since this is an algebraic constraint on the field
imposed at the beginning. 
We shall follow this guideline while computing $\delta'_\xi S'_1$.
Now 
for $S_1$ we normally regard $C^{(4)}$ as independent variable.
If we calculate the corresponding 
change in $S_1$ under general coordinate transformation
of $C^{(4)}$ then in the resulting
expression we would have used the Bianchi identity $dF^{(5)}=0$
already since this is automatic when $F^{(5)}$ is expressed in terms of $C^{(4)}$.
Since under the identification \refb{ek4} this translates to the equation of motion
\refb{e1.5rep} derived from $ S'_1$, 
we can no longer compare $\delta'_\xi S_1$ and $\delta'_\xi S'_1$
off-shell.
To avoid this we 
proceed 
by noting that $S_1$ given in \refb{e4.1}, with the 
identification $\wh F^{(5)}=F^{(5)}+Y$, is invariant 
under general coordinate transformation even if we regard $F^{(5)}$ as an
independent 5-form field variable.  The resulting expression for
$\delta'_\xi S_1$
holds even when $F^{(5)}$ does not satisfy Bianchi identity, and
furthermore 
since this is computed for general $F^{(5)}$, the formula for $\delta'_\xi S_1$ 
will continue to  hold when we impose the self-duality constraint \refb{e4.15} {\it after
computing the variation}. Therefore if
we can ensure that the identity \refb{equal} holds for this expression for 
$\delta'_\xi S_1$, 
this would establish the invariance of $ S'_1$ under
general coordinate transformation.

We begin with the computation of $\delta'_\xi S'_1$. 
It follows from \refb{espirit} and \refb{e1.16rep} that 
\be \label{eyx1}
\delta'_\xi  S'_1 = \int \, d  P^{(4)} \wedge * d \, \delta_\xi P^{(4)} 
-  \int d P^{(4)} \wedge \delta_\xi \, Q^{(5)} -  \int d\, \delta_\xi 
P^{(4)} \wedge Q^{(5)} 
-{1\over 2} \int R^{(5)}\wedge \delta_\xi Q^{(5)}\, .
\ee
This will eventually have to be compared with $\delta_\xi' S_1$. 
Now $\delta_\xi' S_1$
can depend on $Q^{(5)}$ through its dependence on $F^{(5)}$ and the relation
between $F^{(5)}$ and $Q^{(5)}$ encoded in \refb{ek7}, \refb{e2.2rep}, \refb{ek1}, 
but it does not have any
dependence on $P^{(4)}$. Therefore 
the terms involving $P^{(4)}$ in $\delta'_\xi S'_1$
must cancel among themselves. Since we have argued that
neither $\delta_\xi P^{(4)}$ nor
$\delta_\xi Q^{(5)}$ depend on $P^{(4)}$
we see that only the first two terms on the right hand side of \refb{eyx1}
have $P^{(4)}$ dependence and hence they must cancel. This can be achieved by setting
\be \label{eyx2}
\delta_\xi Q^{(5)} = d \, \delta_\xi P^{(4)} + * d \, \delta_\xi P^{(4)}\, .
\ee
With this, \refb{eyx1} reduces to
\be \label{eyx3}
\delta'_\xi  S'_1 =-  \int d\, \delta_\xi 
P^{(4)} \wedge Q^{(5)} 
-{1\over 2} \int R^{(5)}\wedge \delta_\xi Q^{(5)}\, .
\ee
Using \refb{eyx2} again and the anti-self-duality of $R^{(5)}$, we can
express this as
\be \label{eyx4}
\delta'_\xi  S'_1 =-  \int d\, \delta_\xi 
P^{(4)} \wedge (Q^{(5)} - R^{(5)})\, .
\ee

Let us now compute $\delta'_\xi S_1$. 
Using \refb{e4.1} and the fact that
in computing $\delta'_\xi S_1$ we only allow $F^{(5)}$ to vary, we get
\be \label{eyx5}
\delta'_\xi S_1 = - \int \delta_\xi F^{(5)} \wedge (\star_g \wh F^{(5)}-Y)
= -\int \delta_\xi F^{(5)} \wedge F^{(5)}
\, , \ee
where in the second step we have used the self-duality relation \refb{e4.15}  
and the relation $\wh F^{(5)}-Y=F^{(5)}$ given in \refb{e2.2rep}.
Now for any pair of $p$ and $(11-p)$ forms $K^{(p)}$ and
$L^{(11-p)}$ in 9+1 dimensions, we have the general relations
\be \label{epform}
\delta_\xi K^{(p)} = \i_\xi \, d K^{(p)}+ d (\i_\xi \, K^{(p)}), \quad
\int \i_\xi K^{(p)} \wedge L^{(11-p)} = - \int \i_\xi L^{(11-p)}\wedge K^{(p)} \, ,
\ee
where $\i_\xi$ denotes the contraction of $\xi$ with
the differential form. Using this we get
\be \label{eyx6}
\delta_\xi F^{(5)} = d\, \i_\xi F^{(5)} + \i_\xi \, d F^{(5)}\, ,
\ee
and hence
\ben \label{eyx7}
\delta'_\xi S_1 &=& -\int ( d\, \i_\xi F^{(5)} + \i_\xi \, d F^{(5)})
\wedge F^{(5)} =- \int ( d\, \i_\xi F^{(5)} \wedge F^{(5)}
- \i_\xi  F^{(5)} \wedge d F^{(5)}) \nonumber \\
&=& -\int ( d\, \i_\xi F^{(5)} \wedge F^{(5)}
+ d \, \i_\xi  F^{(5)} \wedge  F^{(5)}) =  -2\int  d\, \i_\xi F^{(5)} \wedge F^{(5)}
\, . \een
Comparing the right hand sides of \refb{eyx4} and \refb{eyx7}, and  using
the identification \refb{ek4},
we now see that $\delta'_\xi
S_1$ and $\delta'_\xi   S'_1$ agree if we set
\be \label{eyx8}
\delta_\xi P^{(4)} = \i_\xi F^{(5)} =
\i_\xi (\wh F^{(5)} - Y) \, .
\ee
Using \refb{ek2}, \refb{ek7} this can be expressed as
\be \label{eyx9}
\delta_\xi P^{(4)} = \i_\xi \left[ e\, \zeta\, 
 \left\{ 1 + {1\over 2} (1 +\zeta \ve ) (e\zeta - 1) \right\}^{-1}
\left( {1\over 2} Q^{(5)}  + {1\over 2} 
(1 + \zeta \ve) Y\right) -  Y\right]\, .
\ee
This, together with \refb{eyx2},
determines the general coordinate transformation laws of all the fields appearing
in the action \refb{efin}. 

It is easy to see that when equations of motion are satisfied, the transformation
law of $Q^{(5)}$ given in \refb{eyx2}, \refb{eyx9} agrees with the one induced from the 
transformation law of $F^{(5)}$ via the identification \refb{ek4}. To this end note
that \refb{eyx2} and \refb{eyx8} give
\be \label{estrs}
\delta_\xi Q^{(5)} = d\, \i_\xi F^{(5)} + * d \, \i_\xi F^{(5)}\, .
\ee
On the other hand \refb{ek4} gives
\be \label{eindu}
Q^{(5)} = F^{(5)} + * F^{(5)} \, .
\ee
Thus on-shell, when $dF^{(5)}=0$, the transformation induced on $Q^{(5)}$ from
\refb{eyx6} is
\be \label{eftrs}
\delta_\xi Q^{(5)} = \delta_\xi F^{(5)} + * \delta_\xi F^{(5)}
= d\, \i_\xi F^{(5)} + * d \, \i_\xi  F^{(5)}\, .
\ee
We see that \refb{eftrs} and \refb{estrs} are in perfect agreement.

\refb{eindu} also explains why the transformation laws of $Q^{(5)}$ are
somewhat unusual. Whereas $F^{(5)}$ transforms as a 5-form under general
coordinate transformation, the $*$ in the second term represents Hodge
dual with respect to Minkowski metric, leading to 
non-standard transformation laws of this term.

\sectiono{Supersymmetry} \label{ssuper}

In this section we shall discuss supersymmetry of the action constructed in
\S\ref{s4}.
Our goal will be to propose  supersymmetry
transformation laws $\delta_s'$ 
of the new variables $P^{(4)}$, $Q^{(5)}$ and $M$ that leave the new
action $S_1'+S_2$ given in \refb{etotal}
invariant. We propose the following transformations:
\be \label{esusy}
\delta_s' M = \delta_s M, \quad \delta_s'P^{(4)} = \delta_s C^{(4)}, \quad
\delta_s' Q^{(5)} = 
%\delta_s \left\{ (1 +\zeta \ve ) e\zeta \wt F^{(5)} -
%(1 + \zeta \ve) Y\right\}
d \, \delta_s C^{(4)} + *d \, \delta_s C^{(4)}\, ,
\ee
where $\delta_s$ denotes the usual supersymmetry transformation
laws described
in \cite{schwarz,west2}.
It is understood that on the right hand side of \refb{esusy}
all factors of $dC^{(4)}$ have to
be replaced in terms of $Q^{(5)}$ using \refb{ek2} and \refb{ek7}. To this end
it is important that in the expressions for $\delta_s C^{(4)}$ and $\delta_s M$, 
$C^{(4)}$ always appears in the combination $d C^{(4)}$\cite{west2}, since an explicit
factor of $C^{(4)}$ without derivative 
could not have been expressed back in terms of $Q^{(5)}$. 
Our goal will be to show that $\delta_s'(S_1'+S_2)$ vanishes.
In doing so, we can use
the self-duality of $Q^{(5)}$ since this condition is valid off-shell, 
but not the relation \refb{e1.5rep} since the latter
is an equation of motion derived from $S_1'+S_2$.

For computing $\delta_s'(S_1'+S_2)$ we shall
make use of the known results on the $\delta_s$ transformation properties of the
original action $S_1+S_2$. However instead of regarding $C^{(4)}$ as an independent
variable, it will be more useful for us to regard $F^{(5)}$ as an independent variable
satisfying the self-duality condition $\star_g \wh F^{(5)}=\wh F^{(5)}$. In this case
we can no longer use the Bianchi identity $dF^{(5)}=0$. The expression for
$\delta_s(S_1+S_2)$ under these conditions
can be found using explicit computation, but we shall extract
the result from known results in the literature as follows:
\begin{enumerate}
\item 
An action for type IIB supergravity was proposed in eq.(4.7) of \cite{9806140}.
This action  had, besides the usual fields of type IIB supergravity
which we have called $C^{(4)}$ and
$M$, an additional scalar
field $a$. The scalar field enters the action through a combination $f_4$ which is also
proportional to  $\wh F^{(5)}-\star_g\wh F^{(5)}$. 
The $f_4$ dependent term in the action is quadratic in $f_4$.
We can identify the action $S_1+S_2$ appearing in \refb{es1s2}
as the one given in 
\cite{9806140} without the quadratic term in $f_4$ and without the additional scalar field $a$.
(There are also some obvious changes in the normalizations and notations that can be
easily identified but will not be described here.)
\item The action given  in \cite{9806140} was shown to be invariant under
supersymmetry transformations that agree with those
used in \cite{schwarz,west2} after setting $f_4=0$. 
During this analysis $C^{(4)}$ was taken as the independent variable instead of $F^{(5)}$, and as
a consequence the Bianchi identity $dF^{(5)}=0$ was used.
\item Since the action of \cite{9806140} depends on $f_4$ through a term
quadratic in $f_4$, its first order variation with respect to $f_4$ vanishes at
$f_4=0$. Therefore the supersymmetry of the action of \cite{9806140} guarantees that
the action $S_1+S_2$ is supersymmetric if after taking the supersymmetry variation
we set $\wh F^{(5)}$ to be equal to $\star_g\wh F^{(5)}$ since this sets $f_4$ to 0.
However if we do not use the Bianchi identity $dF^{(5)}=0$ then in general there will
be additional terms proportional to $dF^{(5)}$ in the expression for 
$\delta_s(S_1+S_2)$.
This allows
us to write
\be \label{exs1}
\delta_s(S_1+S_2) = \Xi\, ,
\ee
where $\Xi$ denotes some term proportional to $dF^{(5)}$. 
\item For computing $\Xi$ we can organize each term in 
$\delta_s(S_1+S_2)$ using integration by parts such
that the supersymmetry transformation parameter has no derivative acting on it. In 
this case it is easy to see that since $S_2$ does not depend on $F^{(5)}$,
the entire contribution to $\Xi$ comes from the variation of $S_1$. The variation 
of $C^{(4)}$ generates $\int \delta_s C^{(4)}\wedge d F^{(5)}$. On the other hand,
using the result of \cite{schwarz,west2} that
$\delta Y = - d\, \delta C^{(4)}+\cdots$, where $\cdots$ contain terms
without derivatives of the supersymmetry transformation 
parameters, one finds from
\refb{edy} that
the variation
of $Y$ generates $-2\int \delta_s C^{(4)}\wedge d F^{(5)}$.
This gives
\be \label{eDDelta}
\Xi = -\int  \delta_s C^{(4)}  \wedge  d F^{(5)}  \, .
\ee
\end{enumerate}

Let us now return to our main goal, which is
to show that $\delta_s'(S_1'+S_2)$ vanishes. 
Since $S_2$ depends only
on the set of variables $M$, we have, from \refb{esusy}, $\delta_s' S_2=\delta_s S_2$.
Therefore using \refb{exs1}, we get
\be \label{exs2}
\delta_s'(S_1'+S_2) = \delta_s' S_1' - \delta_s S_1 + \Xi\, .
\ee
The $\Xi$ term on the right hand side is important since using $dF^{(5)}=0$ would
translate to \refb{e1.5rep} under the identification \refb{ek4}, and we are not
allowed to use this relation.
We now note from \refb{espirit}, \refb{e1.16rep}
that
\be \label{esy1}
\delta_s' S_1' = \int \delta_s' P^{(4)}\wedge d \left(Q^{(5)}-* dP^{(4)}\right)+{1\over 2}
\int \delta_s' Q^{(5)} \wedge \, (d P^{(4)} - * d P^{(4)} + R^{(5)})
+ \wt\delta_s  S_1'\, ,
\ee
where $\wt \delta_s$ denotes the variation induced by
$\delta_s'$ (or equivalently $\delta_s$) variation of $M$. 
Using \refb{esusy} this takes the
form
\be \label{esy3}
\delta_s' S_1' = \int \delta_s C^{(4)}\wedge d \left(Q^{(5)}- R^{(5)}\right)
+ \wt\delta_s  S_1'\, .
\ee
On the other hand we have
from \refb{e4.1},
\ben \label{esy2}
\delta_s S_1 &=& -\int \delta_s F^{(5)} \wedge (\star_g \wh F^{(5)} - Y) 
+\wt\delta_s  S_1  = -\int d \, \delta_s C^{(4)} 
 \wedge (\star_g \wh F^{(5)} - Y)  +\wt\delta_s  S_1 \nonumber \\
 &=& \int  \delta_s C^{(4)}  \wedge d
(\wh F^{(5)}-Y) + \wt\delta_s  S_1 =   {1\over 2}  \int \delta_s C^{(4)}\wedge d \left(Q^{(5)}- R^{(5)}\right)
+  \wt\delta_s  S_1\, ,
\een
where in the second line we have used the self-duality constraint $\star_g \wh F^{(5)} 
=\wh F^{(5)} $ which we are allowed to use, and the identification \refb{ek2}, \refb{ek4}.
Now since we have shown in \S\ref{s4} that
the variation of $S_1'$ and $S_1$ with respect to $M$ are identical, 
we have $\wt\delta_s  S_1'=\wt\delta_s  S_1$.
Therefore we get from \refb{exs2}, \refb{esy3}, \refb{esy2}:
\be \label{exs3}
\delta_s'(S_1'+S_2) =  {1\over 2} \int  \delta_s C^{(4)}  \wedge d
 \left(Q^{(5)}- R^{(5)}\right) + \Xi = 0\, ,
\ee
where in the last step we have used
\refb{eDDelta}, \refb{ek2}, \refb{ek4}.
This establishes supersymmetry of the action.

We end the section with two observations:
\begin{enumerate}
\item The form of the
transformation laws given in \refb{esusy}
is consistent with the general form of gauge transformations
described in \cite{1508.05387} and reviewed in \refb{er6},
according to
which the supersymmetry transformation laws of various fields, which is a special
case of the gauge transformation generated by $\lambda$, should be
independent of $P^{(4)}$.
\item
It is easy to verify that the supersymmetry transformation
laws $\delta_s'$ agree with $\delta_s$ after using the identification \refb{ek5}. 
For all fields encoded in $M$ this is automatic consequence of \refb{esusy}; so
we only need to check this for $Q^{(5)}$.  We have from \refb{ek2}, \refb{ek5}
\be
Q^{(5)} = \wh F^{(5)} + * \wh F^{(5)} - Y - * Y = F^{(5)} + * F^{(5)}\, .
\ee
Therefore
\be
\delta_s Q^{(5)} = \delta_s F^{(5)} + * \delta_s F^{(5)} =
d \, \delta_s C^{(4)} + * d \, \delta_s C^{(4)} = \delta_s' Q^{(5)}\, ,
\ee
where in the last step we have used \refb{esusy}.
This shows that the transformations $\delta_s$ and $\delta_s'$ agree.
\end{enumerate}

\sectiono{Lorentz covariant gauge fixing and Feynman rules} \label{s6.5}

String field theory action of \cite{1508.05387} admits a Lorentz 
covariant gauge fixing at the
full quantum level -- the `Siegel gauge'.  This suggests that the action
given in \refb{efin} (together with $S_2$) 
must also admit a Lorentz covariant gauge fixing.
In this section we shall describe how this can be done 
in flat space-time background. 

Since gauge transformations of most fields are
standard and we can choose the analog of Lorentz / Feynman gauge for them
maintaining manifest Lorentz covariance, we shall focus on the 
$P^{(4)} \to P^{(4)}+d\, \Xi^{(3)}$ gauge transformation. We can fix a gauge 
by adding
a gauge fixing term of the form
\be  \label{egaugefix}
{1\over 2} \int * \, d * P^{(4)} \wedge \, d * P^{(4)} \, .
\ee
Since in flat space-time 
the background value of $e$ is $\eta$, $(e\eta-1)$ and hence 
$\MM$ has its expansion beginning at the first order in the fluctuations. Therefore
the only terms quadratic in $P^{(4)}$, $Q^{(5)}$ in the original action
are the first two terms on the right
hand side of \refb{efin}. 
After adding \refb{egaugefix} 
to the action \refb{efin} the quadratic term involving 
$P^{(4)}$ and $Q^{(5)}$ takes the form
\be 
{1\over 2} \int P^{(4)} \wedge *(*\, d*d + d*d\, *) P^{(4)} 
+  \int P^{(4)}\wedge d Q^{(5)}\, .
\ee
In momentum space this corresponds to a term proportional to
\ben \label{ekinetic}
&& {1\over 2\, \times 4!} \int d^{10} k \left[P^{(4)abcd} (-k)
\, k^2 \, P^{(4)}_{abcd}(k) 
+2\,  i\, P^{(4)abcd} (-k)  k^e Q^{(5)}_{eabcd}(k)\right] \nonumber \\
&=& {1\over 2\, \times 4!} \int d^{10} k \bigg[(P^{(4)abcd} (-k) 
- i (k^2)^{-1} k_f Q^{(5)fabcd}(-k) )
\, k^2 \, (P^{(4)}_{abcd}(k) + i (k^2)^{-1} k^e Q^{(5)}_{eabcd}(k)) 
\nonumber \\
&&  - Q^{(5)fabcd}(-k) (k^2)^{-1}  k_f \, k^e Q^{(5)}_{eabcd}(k) 
\bigg] \nonumber \\ 
&=& {1\over 2\, \times 4!} \int d^{10} k \left[\bar P^{(4)abcd} (-k)
\, k^2 \, \bar P^{(4)}_{abcd}(k) -  Q^{(5)fabcd}(-k) (k^2)^{-1}  k_f \, k^e 
Q^{(5)}_{eabcd}(k) 
\right]\, ,
\een
where
\be 
\bar P^{(4)}_{abcd}(k) \equiv P^{(4)}_{abcd}(k) + i (k^2)^{-1} k^e 
Q^{(5)}_{eabcd}(k)\, .
\ee
We can now 
treat $\bar P^{(4)}$ as the independent field instead of $P^{(4)}$. Since
this does not appear anywhere else in the action, 
this describes a free field and hence decouples. Therefore the only 
kinetic operator that is of relevance is that of $Q^{(5)}$. 
If we define the following operator acting on 5-forms:
\be
K_A{}^B (k) = (k^2)^{-1} \left(
k_{a_1} k^{b_1} \delta_{a_2}{}^{b_2} \cdots \delta_{a_5}{}^{b_5}
+ \delta_{a_1}{}^{b_1} k_{a_2} k^{b_2} \delta_{a_3}{}^{b_3} \delta_{a_4}{}^{b_4}
\delta_{a_5}{}^{b_5} + \cdots +  \delta_{a_1}{}^{b_1} \cdots  \delta_{a_4}{}^{b_4}
k_{a_5} k^{b_5}\right)\, ,
\ee
then the kinetic operator acting on $Q^{(5)}$ may be written as
\be \label{eKin}
-{1\over 4}\, \zeta \, (1-\zeta\ve) \, K(k) \, (1+\zeta\ve)\, ,
\ee
up to a constant of proportionality.
The operator $(1+\zeta\ve)/2$ on the right
projects onto self-dual 5-forms, whereas the
operator $(1-\zeta\ve)/2$ on the left projects onto anti-self-dual 5-forms 
reflecting the fact that only anti-self-dual 5-forms have non-zero
contraction with self-dual 5-forms. 
Thus the kinetic operator is a map from the space of self-dual 5-forms to the space of
anti-self-dual 5-forms.
The propagator, which is $i$ 
times the inverse of
the kinetic term, should be a map from the space of anti-self-dual 5-forms
to the space of self-dual 5-forms, also reflecting the fact that the propagator
naturally acts on current dual to field which is in this case anti-self-dual
5-form. It is easy to verify that the following operator constitutes the
inverse of the kinetic term in this sense:
\be \label{eDelta}
\Delta =  -(1+\zeta\ve) \, K(k) \, (1-\zeta\ve)\, \zeta\, .
\ee
This is the gauge invariant propagator of a 5-form field strength given {\it e.g.}
in \cite{alvarez}.
With this propagator $i\Delta$ 
for the $Q^{(5)}$ field, and the vertices computed in the
usual way from the action $S_1'+S_2$, we can now compute the tree level
Green's functions and S-matrix elements of type IIB supergravity in the 
standard way. Loop corrections will require embedding this theory into the
full string field theory described in \cite{1508.05387}.

\bigskip

\noindent {\bf Acknowledgement:}
I wish to thank Nathan Berkovits, Yuji Okawa, Martin Schnabl,
John Schwarz and
Barton Zwiebach for valuable discussions.
This work  was
supported in part by the 
DAE project 12-R\&D-HRI-5.02-0303 and J. C. Bose fellowship of 
the Department of Science and Technology, India.

\end{document}